\documentclass[aps,preprint]{revtex4}%
\usepackage{amssymb}
\usepackage{amsfonts}
\usepackage{amsmath}
\usepackage{graphicx}
\usepackage[usenames,dvipsnames]{color}%
\providecommand{\U}[1]{\protect\rule{.1in}{.1in}}

\begin{document}
\preprint{ }
\title{Analytic method for quadratic polarons in nonparabolic bands}
\author{S. N. Klimin}
\author{J. Tempere}
\thanks{Also at Lyman Laboratory of Physics, Harvard University, Cambridge, MA 02138, USA.}
\author{M. Houtput}
\affiliation{TQC, Departement Fysica, Universiteit Antwerpen, Universiteitsplein 1, B-2610
Antwerpen, Belgium}
\author{S. Ragni}
\affiliation{Faculty of Physics, Computational Materials Physics, University of Vienna,
Kolingasse 14-16, Vienna A-1090, Austria}
\author{T. Hahn}
\thanks{Center for Computational Quantum Physics, Flatiron Institute, 162 5th Avenue,
New York, New York 10010, USA}
\author{C. Franchini}
\thanks{Also at Department of Physics and Astronomy \textquotedblleft Augusto
Righi\textquotedblright, Alma Mater Studiorum -- Universit\`{a} di Bologna,
Bologna, 40127 Italy}
\affiliation{Faculty of Physics, Computational Materials Physics, University of Vienna,
Kolingasse 14-16, Vienna A-1090, Austria}
\author{A. S. Mishchenko}
\thanks{Also at RIKEN Center for Emergent Matter Science (CEMS), Wako, Saitama
351-0198, Japan}
\affiliation{Department for Research of Materials under Extreme Conditions, Institute of
Physics, 10000 Zagreb, Croatia}

\begin{abstract}
Including the effect of lattice anharmonicity on electron-phonon interactions
has recently garnered attention due to its role as a necessary and significant
component in explaining various phenomena, including superconductivity,
optical response, and temperature dependence of mobility. This study focuses
on analytically treating the effects of anharmonic electron-phonon coupling on
the polaron self-energy, combined with numerical Diagrammatic Monte Carlo
data. Specifically, we incorporate a quadratic interaction into the method of
squeezed phonon states, which has proven effective for analytically
calculating the polaron parameters. Additionally, we extend this method to
nonparabolic finite-width conduction bands while maintaining the periodic
translation symmetry of the system. Our results are compared with those
obtained from Diagrammatic Monte Carlo, partially reported in a recent study
[S. Ragni \emph{et al}., \emph{Phys.~Rev.~B \textbf{107}, L121109}%
(\emph{2023})], covering a wide range of coupling strengths for the nonlinear
interaction. Remarkably, our analytic method predicts the same features as the
Diagrammatic Monte Carlo simulation.

\end{abstract}
\date{\today}
\maketitle

\section{Introduction \label{Sec:Intro}}

Polaron physics, which originates from a theoretical problem that involves the
interaction of a particle with a quantum field \cite{Feynman,DevreeseBook},
has garnered significant experimental interest due to its practical
applications. This curiosity has additionally driven the advancement of
polaron theory, which has broadened its scope from its original concentration
on polarons in crystals~\cite{franchini2021} to include a range of condensed
matter systems, such as quantum gases \cite{Schirotzek, Kohstall,
Shchadilova}, and even celestial bodies like neutron stars
\cite{Kutschera1993, Nascimbene2010}.

Conventional theoretical frameworks for polarons typically center around the
idea of small oscillations within a crystal lattice or another bosonic quantum
field. Within this framework, the linear-harmonic approximation characterizes
the phonon field as harmonic, with the electron-phonon interaction being
directly proportional to the phonon coordinates. Instances of such frameworks
include the Fr\"{o}hlich and Holstein models for polarons in solid-state
physics \cite{DevreeseBook}, as well as Fr\"{o}hlich-type Hamiltonians for
impurity polarons in quantum gases \cite{Tempere2009}. In these applications,
the Fr\"{o}hlich model relies on the assumption of a parabolic energy
dispersion for the conduction band.

It has been established for a while that in certain situations, higher-order
terms beyond the linear-harmonic approximation can be significant
\cite{Epifanov}. These additional terms specifically influence the optical
response and kinetics of impurities within crystals. Lately, there has been a
resurgence of interest in nonlinear electron-phonon interactions and
anharmonic phonons
\cite{Kussow,Errea,Maslov,Houtput2021,PRBL2023,Adolphs,Ranalli2023,
Verdi2023}. These occurrences have proven crucial in elucidating different
phenomena, such as superconductivity at low carrier concentrations
\cite{VDMarel2,Gastiasoro2020} and the temperature-dependent mobility
\cite{Schilcher2021}.

While Diagrammatic Monte Carlo (DiagMC) simulations have made it possible to
describe polaron properties for both linear \cite{Mishchenko2000, Hahn2018}
and nonlinear \cite{PRBL2023} electron-phonon interactions with high accuracy,
analytic methods remain of significant interest. These methods provide a clear
physical picture of polarons and enhance our understanding of results obtained
through numerical techniques such as density functional theory and DiagMC simulations.

The primary objective of the present study is to develop an analytical method
for investigating anharmonic polarons within a nonparabolic conduction band
characterized by a finite bandwidth. By adjusting the bandwidth and exploring
various electron-phonon interaction amplitudes, this framework encompasses
scenarios involving both small and large polarons, as well as the intermediate
regime between these extremes. Our approach is based on the well-established
method of using displaced squeezed phonon states \cite{Gross,Kandemir} after
elimination of the electron coordinate by a shift to the frame co-moving with
the electron \cite{LLP}. This technique has been widely used for large
polarons with linear interactions in weak and intermediate coupling regimes.

One of the important advances in the understanding of the role of high-order
terms in the electron-phonon coupling is presented in \cite{Adolphs} where the
semi-analytic method of momentum average approximation is used to study the
Holstein model with simultaneous quadratic and quartic terms. Unfortunately,
we cannot verify our results against the data in \cite{Adolphs} because we
consider a different model, with only quadratic interaction whereas the
authors of \cite{Adolphs} do not provide results where the quartic term is
missing. Also, the paper \cite{Adolphs} deals with a polaron in 1D and 2D,
while the present work is devoted to a polaron in three dimensions.

In this work, we first extend the displaced squeezed phonon approximation to
incorporate quadratic interactions (discussed in Sec.~\ref{Sec:2TO}). Next, in
Sec.~\ref{Sec:Scalar}, we examine polaron behavior with a quadratic
interaction in nonparabolic bands while maintaining periodic boundary
conditions for the Brillouin zone. This polaron model has been recently
explored numerically \cite{PRBL2023} using the DiagMC technique, and we
compare our results with those of DiagMC. Finally, the results obtained are
summarized in Sec.~\ref{sec:Conclusions}.

\section{Polaron with Fr\"{o}hlich and 2TO interactions \label{Sec:2TO}}

\subsection{The system}

In this section, we investigate the polaron in a parabolic conduction band,
focusing on two distinct types of electron-phonon interactions. First, we
examine the Fr\"{o}hlich interaction, which involves longitudinal optical (LO)
phonons and exhibits a linear dependence on phonon coordinates. Second, we
explore the quadratic interaction which engages two transverse optical (TO)
phonons. This 2TO interaction manifests itself as a quadratic function of
phonon coordinates. Although the 2TO interaction was introduced in theoretical
frameworks long ago \cite{Epifanov}, it has recently regained interest.
Notably, a coexistence of Fr\"{o}hlich and 2TO couplings has proven relevant
in experimental contexts, particularly with materials such as SrTiO$_{3}$
\cite{VDMarel2,Gastiasoro2020, Verdi2023}. These interactions have allowed us
to successfully elucidate various phenomena, including response properties and superconductivity.

Our analysis centers around the electron-phonon Hamiltonian:%
\begin{align}
H &  =\frac{\mathbf{p}^{2}}{2m}+\sum_{a=1}^{3}\sum_{\mathbf{q}}\hbar
\omega_{\mathbf{q}}^{\left(  a\right)  }b_{\mathbf{q}}^{\left(  a\right)
\dag}b_{\mathbf{q}}^{\left(  a\right)  }\nonumber\\
&  +\sum_{\mathbf{q}}\left(  V_{\mathbf{q}}b_{\mathbf{q}}^{\left(  3\right)
}e^{i\mathbf{q\cdot r}}+V_{\mathbf{q}}^{\ast}b_{\mathbf{q}}^{\dag\left(
3\right)  }e^{-i\mathbf{q\cdot r}}\right)  +H_{2\mathrm{TO}}\label{Ham}%
\end{align}
Here, $\mathbf{r}$ is the position operator of the electron with band mass
$m$, $\mathbf{p}$ is its canonically conjugate momentum operator, and
$b_{\mathbf{q}}^{\dagger\left(  a\right)  }$ and $b_{\mathbf{q}}^{\left(
a\right)  }$ are the creation and annihilation operators for optical phonons
of wave vector $\mathbf{q}$ and energy $\hbar\omega_{\mathbf{q}}^{\left(
a\right)  }$. The index $a=1,2$ labels the TO phonon modes, and $a=3$ denotes
the LO mode. So further on in this note we assume that $\omega_{\mathbf{q}%
}^{\left(  1\right)  }=\omega_{\mathbf{q}}^{\left(  2\right)  }=\omega
_{\mathbf{q}}^{\left(  {\text{TO}}\right)  }$ and $\omega_{\mathbf{q}%
}^{\left(  3\right)  }=\omega_{\text{LO}}$. The $V_{\mathbf{q}}$ are Fourier
components of the linear (Fr\"{o}hlich) part of the electron-phonon
interaction
\begin{equation}
V_{\mathbf{q}}=\frac{\hbar\omega_{\text{LO}}}{q}\left(  \frac{4\pi\alpha}%
{V}\right)  ^{\frac{1}{2}}\left(  \frac{\hbar}{2m\omega_{\text{LO}}}\right)
^{\frac{1}{4}}.\label{eq_1b}%
\end{equation}
The strength of the Fr\"{o}hlich electron--phonon interaction is {expressed
by} a dimensionless coupling constant $\alpha$, which is defined as:
\begin{equation}
\alpha=\frac{e^{2}}{\hbar}\sqrt{\frac{m}{2\hbar\omega_{\text{LO}}}}\left(
\frac{1}{\varepsilon_{\infty}}-\frac{1}{\varepsilon_{0}}\right)
.\label{eq_1c}%
\end{equation}
In this definition, $\varepsilon_{\infty}$ and $\varepsilon_{0}$ are,
respectively, the high-frequency and static dielectric constants of the polar crystal.

We apply the quadratic electron-phonon interaction Hamiltonian following
Ref.~\cite{Maslov}:%
\begin{equation}
H_{2\mathrm{TO}}=\frac{g_{2}}{2}\vec{P}^{2}\left(  \mathbf{r}\right)
\label{H2TO1}%
\end{equation}
where $g_{2}$ is the coupling strength for the 2TO interaction. The
polarization $\vec{P}\left(  \mathbf{r}\right)  $ is given by%
\begin{equation}
\vec{P}\left(  \mathbf{r}\right)  =\sum_{\mathbf{q}}\sum_{a=1,2}%
\frac{\mathbf{e}_{\mathbf{q}}^{\left(  a\right)  }}{\sqrt{V}}\varkappa
_{\mathbf{q}}\left(  b_{\mathbf{q}}^{\left(  a\right)  }e^{i\mathbf{q}%
\cdot\mathbf{r}}+b_{\mathbf{q}}^{\left(  a\right)  \dag}e^{-i\mathbf{q}%
\cdot\mathbf{r}}\right)  \label{Pr}%
\end{equation}
with the factor to the interaction strength,%
\begin{equation}
\varkappa_{\mathbf{q}}=\sqrt{\omega_{\mathbf{q}}^{\left(  {\text{TO}}\right)
}\frac{\varepsilon_{0}\left(  \mathbf{q}\right)  -\varepsilon_{\infty}}{4\pi}}
\label{Aq}%
\end{equation}
where $\varepsilon_{0}\left(  \mathbf{q}\right)  $ and $\varepsilon_{\infty}$
are, respectively, the static momentum-dependent dielectric function and
high-frequency dielectric constant. In the considered isotropic model, there
are two unit vectors $\mathbf{e}_{\mathbf{q}}^{a}$ for TO modes, with $a=1,2$,
orthogonal to $\mathbf{q}$ and to each other.

\subsection{The approximation of squeezed phonon states}

The original idea of the Lee-Low-Pines{ transformation} \cite{LLP} followed by
the Bogoliubov-Tyablikov diagonalization \cite{Tyablikov} of the truncated
coordinate-free polaron Hamiltonian, quadratic in phonon coordinates, belongs
to Gross \cite{Gross}. This approximation was subsequently further developed
in different modifications, e.g., the method of displaced squeezed phonon
states \cite{Kandemir} or correlated Gaussian wave functions
\cite{Shchadilova} (see also Refs.~\cite{Tulub1962,Tulub2015,Porsch}). Here,
we show that the approximation of displaced squeezed phonon states
(abbreviated to SPS in the figures) is straightforwardly extended to a polaron
with a quadratic electron-phonon interaction.

The first Lee-Low-Pines unitary transformation%
\begin{equation}
S_{1}=\exp\left[  \frac{i}{\hbar}(\mathbf{P}-\sum_{a,\mathbf{q}}%
\hbar\mathbf{q}b_{\mathbf{q}}^{\left(  a\right)  \dag}b_{\mathbf{q}}^{\left(
a\right)  })\mathbf{\cdot r}\right]  , \label{S1}%
\end{equation}
where $\mathbf{P}$ is the eigenvalue of the total momentum, leads to the
coordinate-free polaron Hamiltonian%
\begin{align}
\mathcal{H}  &  =S_{1}^{-1}HS_{1}\nonumber\\
&  =\frac{\left(  \mathbf{P}-\sum_{a,\mathbf{q}}\hbar\mathbf{q}b_{\mathbf{q}%
}^{\left(  a\right)  \dag}b_{\mathbf{q}}^{\left(  a\right)  }\right)  ^{2}%
}{2m}+\sum_{a=1}^{3}\sum_{\mathbf{q}}\hbar\omega_{\mathbf{q}}^{\left(
a\right)  }b_{\mathbf{q}}^{\left(  a\right)  \dag}b_{\mathbf{q}}^{\left(
a\right)  }\nonumber\\
&  +\sum_{\mathbf{q}}V_{\mathbf{q}}B_{\mathbf{q}}^{\left(  3\right)  }%
+\frac{g_{2}}{2V}\left(  \sum_{\mathbf{q},a}\varkappa_{\mathbf{q}}%
\mathbf{e}_{\mathbf{q}}^{\left(  a\right)  }B_{\mathbf{q}}^{\left(  a\right)
}\right)  ^{2}. \label{HCF}%
\end{align}
Here $B_{\mathbf{q}}^{\left(  a\right)  }$ is proportional to the phonon
coordinate, and is given by
\begin{equation}
B_{\mathbf{q}}^{\left(  a\right)  }=b_{\mathbf{q}}^{\left(  a\right)
}+b_{\mathbf{q}}^{\left(  a\right)  \dag}. \label{Bk}%
\end{equation}

The coordinate-free Hamiltonian looks appealing for approximate methods and
led to numerous attempts to make analytic approximations suitable at both weak
and strong coupling. Here, we consider the well-established approach for the
weak- and intermediate-coupling regime. In this approach, the second
Lee-Low-Pines transformation is performed,
\begin{equation}
S_{2}=\exp\left[  -\sum_{a,\mathbf{q}}f_{\mathbf{q}}^{\left(  a\right)
}\left(  b_{\mathbf{q}}^{\left(  a\right)  }-b_{\mathbf{q}}^{\left(  a\right)
\dagger}\right)  \right]  . \label{LLP2}%
\end{equation}
The phonon shifts $f_{\mathbf{q}}^{\left(  a\right)  }$ are real, in
accordance with the polaron Hamiltonian. The unitary transformation $S_{2}$
results in a displacement of the phonon operators. The resulting transformed
Hamiltonian consists of two terms:%
\[
S_{2}^{-1}\mathcal{HS}_{2}=H_{0}+H_{I}%
\]
where $H_{0}$ and $H_{I}$ are the following contributions to the
coordinate-free Hamiltonian after phonon shifts:\newline(1) The Hamiltonian
truncated to the quadratic normal form of phonon operators:%
\begin{align}
H_{0}  &  =E_{0}+\sum_{a,\mathbf{q}}\hbar\Omega_{\mathbf{q}}^{\left(
a\right)  }b_{\mathbf{q}}^{\left(  a\right)  \dag}b_{\mathbf{q}}^{\left(
a\right)  }\nonumber\\
&  +\frac{1}{2m}\left(  \sum_{a,\mathbf{q}}\hbar\mathbf{q}f_{\mathbf{q}%
}^{\left(  a\right)  }B_{\mathbf{q}}^{\left(  a\right)  }\right)  ^{2}%
+\frac{g_{2}}{2V}\left(  \sum_{\mathbf{q},a=1,2}\mathbf{e}_{\mathbf{q}%
}^{\left(  a\right)  }\varkappa_{\mathbf{q}}B_{\mathbf{q}}^{\left(  a\right)
}\right)  ^{2} \label{H0}%
\end{align}
with the renormalized phonon energy,%
\begin{equation}
\hbar\Omega_{\mathbf{q}}^{\left(  a\right)  }=\hbar\omega_{\mathbf{q}%
}^{\left(  a\right)  }+\frac{\hbar^{2}\mathbf{q}^{2}}{2m}-\frac{\hbar\left(
\mathbf{q}\cdot\mathbf{P}\right)  }{m}+\sum_{a^{\prime},\mathbf{q}^{\prime}%
}\frac{\hbar^{2}}{m}\left(  \mathbf{q}\cdot\mathbf{q}^{\prime}\right)  \left(
f_{\mathbf{q}^{\prime}}^{\left(  a^{\prime}\right)  }\right)  ^{2} \label{Wka}%
\end{equation}
where the term $E_{0}$ does not contain operators:%
\begin{align}
E_{0}  &  =\frac{\mathbf{P}^{2}}{2m}+\sum_{a,\mathbf{q}}\left(  \hbar
\omega_{\mathbf{q}}^{\left(  a\right)  }-\frac{\left(  \hbar\mathbf{q}%
\cdot\mathbf{P}\right)  }{m}\right)  \left(  f_{\mathbf{q}}^{\left(  a\right)
}\right)  ^{2}+2\sum_{\mathbf{q}}V_{\mathbf{q}}f_{\mathbf{q}}^{\left(
3\right)  }\nonumber\\
&  +\frac{1}{2m}\left(  \sum_{a,\mathbf{q}}\hbar\mathbf{q}\left(
f_{\mathbf{q}}^{\left(  a\right)  }\right)  ^{2}\right)  ^{2}+\frac{2g_{2}}%
{V}\left(  \sum_{\mathbf{q},a=1,2}\mathbf{e}_{\mathbf{q}}^{\left(  a\right)
}\varkappa_{\mathbf{q}}f_{\mathbf{q}}^{\left(  a\right)  }\right)  ^{2},
\label{E0}%
\end{align}
(2) The remaining part of the Hamiltonian, which is also written in the normal
form:%
\begin{align}
H_{I}  &  =\sum_{\mathbf{q}}\left(  \hbar\Omega_{\mathbf{q}}^{\left(
3\right)  }f_{\mathbf{q}}^{\left(  3\right)  }+V_{\mathbf{q}}\right)
B_{\mathbf{q}}^{\left(  3\right)  }\nonumber\\
&  +\sum_{\mathbf{q},a=1,2}\left(  \hbar\Omega_{\mathbf{q}}^{\left(  a\right)
}f_{\mathbf{q}}^{\left(  a\right)  }+\frac{2g_{2}}{V}\sum_{\mathbf{q}^{\prime
},a^{\prime}=1,2}\left(  \mathbf{e}_{\mathbf{q}}^{\left(  a\right)  }%
\cdot\mathbf{e}_{\mathbf{q}^{\prime}}^{\left(  a^{\prime}\right)  }\right)
\varkappa_{\mathbf{q}}\varkappa_{\mathbf{q}^{\prime}}f_{\mathbf{q}^{\prime}%
}^{\left(  a^{\prime}\right)  }\right)  B_{\mathbf{q}}^{\left(  a\right)
}\nonumber\\
&  +\sum_{a,\mathbf{q}}\sum_{a^{\prime},\mathbf{q}^{\prime}}\frac{\hbar
^{2}\left(  \mathbf{q}\cdot\mathbf{q}^{\prime}\right)  }{m}f_{\mathbf{q}%
}^{\left(  a\right)  }b_{\mathbf{q}^{\prime}}^{\left(  a^{\prime}\right)
\dag}B_{\mathbf{q}}^{\left(  a\right)  }b_{\mathbf{q}^{\prime}}^{\left(
a^{\prime}\right)  }\nonumber\\
&  +\sum_{a,\mathbf{q}}\sum_{a^{\prime},\mathbf{q}^{\prime}}\frac{\hbar
^{2}\left(  \mathbf{q}\cdot\mathbf{q}^{\prime}\right)  }{2m}b_{\mathbf{q}%
}^{\left(  a\right)  \dag}b_{\mathbf{q}^{\prime}}^{\left(  a^{\prime}\right)
\dag}b_{\mathbf{q}}^{\left(  a\right)  }b_{\mathbf{q}^{\prime}}^{\left(
a^{\prime}\right)  }. \label{HI}%
\end{align}
The term $H_{I}$ can only quantitatively influence the polaron energy in a
sufficiently strong coupling. Consequently, an account of this contribution is
beyond the scope of the present work, which is restricted to weak- and
intermediate-coupling regimes.

\subsection{2TO polaron self-energy}

The Hamiltonian $H_{0}$, given by Eq.~(\ref{H0}), is quadratic in the phonon
operators, but it does contain products of two creation or two annihilation
operators. Such Hamiltonians can be diagonalized by a Bogoliubov
transformation, which can be interpreted as a unitary transformation
representing a phonon \emph{squeezing}. We refer to this transformation as the
Bogoliubov-Tyablikov diagonalization \cite{Tyablikov}. The momentum-dependent
polaron energy shift provided by the Bogoliubov-Tyablikov diagonalization is
found starting from the definition \cite{Wentzel}%
\begin{equation}
\Delta E=\frac{1}{2}\sum_{a,\mathbf{q}}\hbar\left(  \nu_{\mathbf{q}}^{\left(
a\right)  }-\Omega_{\mathbf{q}}^{\left(  a\right)  }\right)  , \label{DE}%
\end{equation}
where $\nu_{\mathbf{q}}^{\left(  a\right)  }$ are eigenfrequencies determined
in Appendix A which satisfy Eq. (\ref{eigenf}). In fact, the energy $\Delta E$
is calculated exactly without the necessity to know the eigenfrequencies
explicitly -- the details are described in Appendix A. The resulting polaron
self-energy consists of (\ref{E0}) and the self-energy correction due to the
squeezing transformation,%
\begin{equation}
E_{p}\left(  \mathbf{P}\right)  =E_{0}+\Delta E. \label{Epol}%
\end{equation}

\begin{figure}[ptbh]%
\centering
\includegraphics[
height=1.6527in,
width=3.2863in
]%
{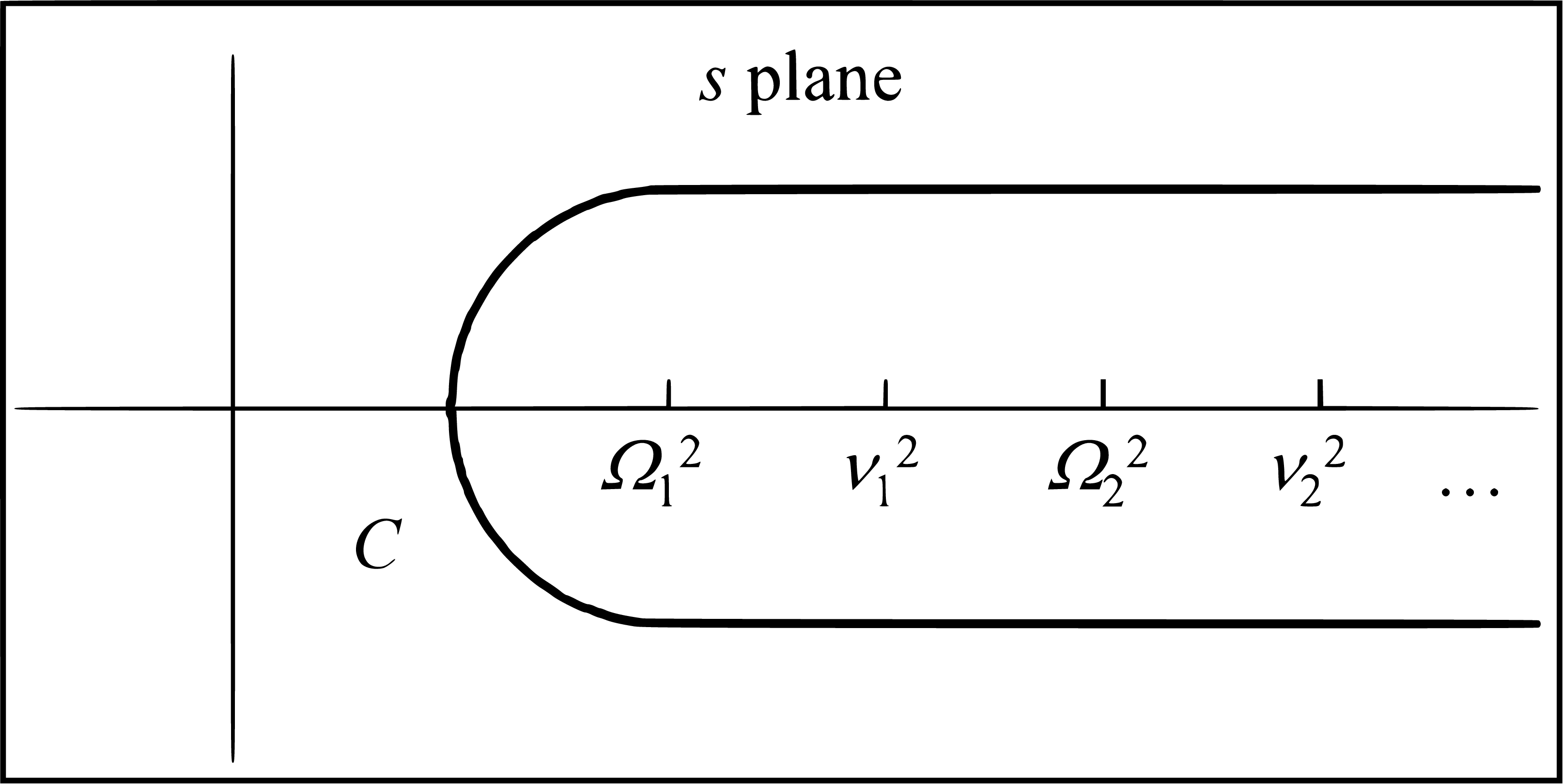}%
\caption{Integration contour in the complex $s$ plane for the polaron energy
shift.}%
\label{fig:Contour}%
\end{figure}

Polarons of different types with linear electron-phonon coupling within the
displaced squeezed phonon approximation were extensively studied in earlier
work, e.g. \cite{Gross,Tulub1962,Tulub2015,Shchadilova}. Therefore, here we
focus only on the self-energy of the 2TO polaron. To understand the effects of
quadratic coupling more clearly, we restrict the results analysis to the case
of a purely quadratic polaron without linear electron-phonon interaction. In
general, when both linear and quadratic interactions are present in the
Hamiltonian, $f_{\mathbf{q}}^{\left(  a\right)  }$ are treated as variational
functions and are chosen to minimize the polaron self-energy (\ref{Epol}). In
the absence of linear coupling, the optimal phonon shift values appear to be
zero. Thus, the polaron energy shift for a purely quadratic polaron is only
expressed by the term $\Delta E$. It is derived using the known scheme of the
meson pair theory \cite{Wentzel} and expressed through the contour integral
\begin{equation}
\Delta E^{\left(  2\mathrm{TO}\right)  }\left(  P\right)  =-\frac{\hbar}{8\pi
i}\oint_{C}ds~\frac{1}{\sqrt{s}}\ln\Delta^{\left(  2\mathrm{TO}\right)
}\left(  s,P\right)  \label{DE2TO}%
\end{equation}
with the function (see Appendix A)%
\begin{equation}
\ln\Delta^{\left(  2\mathrm{TO}\right)  }\left(  s,P\right)  =\sum
_{j=x,y,z}\ln\left[  1-\frac{g_{2}}{\hbar V}\sum_{\mathbf{q},a=1,2}\left(
1-\frac{q_{j}^{2}}{q^{2}}\right)  \frac{\varkappa_{\mathbf{q}}^{2}%
\Omega_{\mathbf{q}}^{\left(  a\right)  }}{s-\left(  \Omega_{\mathbf{q}%
}^{\left(  a\right)  }\right)  ^{2}}\right]  . \label{log}%
\end{equation}
The integration contour $C$ contains inside all values of $\left(
\nu_{\mathbf{q}}^{\left(  a\right)  }\right)  ^{2}$ and $\left(
\Omega_{\mathbf{q}}^{\left(  a\right)  }\right)  ^{2}$ as shown in Fig.
\ref{fig:Contour}.%

Here, for numeric testing, we use the approximation of a soft TO mode applied
by Kumar \emph{et al}. \cite{Maslov}:
\[
\omega_{\mathbf{q}}^{\left(  {\text{TO}}\right)  }=\sqrt{\omega_{T}^{2}%
+c^{2}q^{2}}%
\]
and neglect $\left.  \omega_{\mathbf{q}}^{\left(  {\text{TO}}\right)
}\right\vert _{q=0}=\omega_{T}$ with respect to $cq$. Thus the soft mode is
approximately sound-like:
\begin{equation}
\omega_{\mathbf{q}}^{\left(  {\text{TO}}\right)  }\approx cq,\qquad
\varkappa_{\mathbf{q}}^{2}\approx\frac{\varepsilon_{\infty}\omega_{\text{LO}%
}^{2}}{4\pi cq} \label{wq}%
\end{equation}

Without knowledge of a realistic large-$q$ behavior of the TO-phonon energy
and of the 2TO interaction, the integrals over the phonon momentum in the
present approximation diverge. To remove this divergence a phonon wave vector
cutoff $k_{0}$ is introduced. The coupling strength of the 2TO interaction is
expressed through the dimensionless coupling constant%
\begin{equation}
\alpha_{T}=g_{2}\frac{\varepsilon_{\infty}m\omega_{\text{LO}}^{2}}{6\pi
^{3}\hbar^{2}c}. \label{alphaT}%
\end{equation}
%

\begin{figure}[ptbh]%
\centering
\includegraphics[
height=3.1929in,
width=3.9159in
]%
{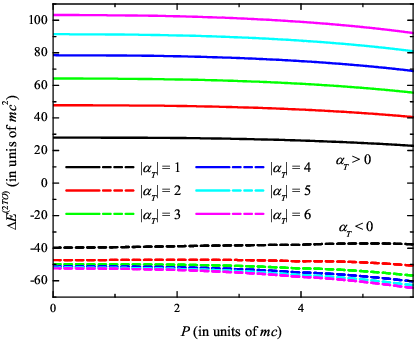}%
\caption{Momentum-dependent TO2-polaron energy shift as a function of the
polaron momentum $P$ for different values of the coupling constant $\alpha
_{T}$, calculated using the momentum cutoff $\hbar k_{0}=10mc$. Solid and
dashed curves show the energy for positive and negative $\alpha_{T}$,
respectively.}%
\label{fig:enrK10}%
\end{figure}

The polaron energy shift (\ref{DE2TO}) describes the dispersion of the polaron
$\Delta E^{\left(  2\mathrm{TO}\right)  }\left(  P\right)  $ as a function of
the polaron momentum $P$. The numeric results for this 2TO contribution to the
polaron self-energy are shown in Fig. \ref{fig:enrK10}. The polaron energies
$\Delta E^{\left(  2\mathrm{TO}\right)  }$ are plotted as functions of the
total momentum $P$ for different values of the coupling constant $\alpha_{T}$.
The numeric calculation is performed using the units with $\hbar=1$, $m=1$,
and $c=1$. Thus the energy is measured in units of $mc^{2}$.

To our knowledge, the sign of the coupling constant for the 2TO interaction is
not known \emph{a priori}. Therefore, the polaron self-energy is calculated
here for both positive and negative $\alpha_{T}$, shown by solid and dashed
curves, respectively. The range of the polaron momentum is chosen sufficiently
small with respect to the momentum cutoff in order to avoid possible artifacts
related to the cutoff. As can be seen in Fig. \ref{fig:enrK10}, the 2TO
polaron energy shift $\Delta E^{\left(  2\mathrm{TO}\right)  }\left(
P\right)  $ smoothly and monotonically decreases as a function of $P$.%

\begin{figure}[ptbh]%
\centering
\includegraphics[
height=3.5734in,
width=5.0323in
]%
{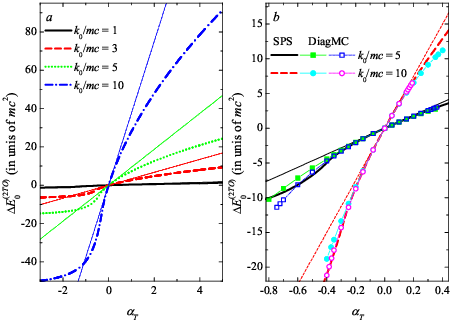}%
\caption{(\emph{a}) TO2-polaron ground state energy calculated using the
approximations for the soft-mode TO-phonon dispersion and the coupling factor
by Kumar \emph{et al}. \cite{Maslov} for different values of the phonon cutoff
momentum $k_{0}$. The thin lines show the first-order perturbation results for
$E^{\left(  2TO\right)  }$. (\emph{b}) The ground state polaron energy within
the approximation of squeezed phonon states compared with results of partial
summation of DiagMC series containing up to 2-loop diagrams (filled symbols)
and up to 3-loop diagrams (hollow symbols). Errorbars of DiagMC data are
smaller than the points size.}%
\label{fig:gsenergy}%
\end{figure}

Expanding the momentum-dependent 2TO polaron energy shift (\ref{DE2TO}) in
powers of $P$ up to the second order, we obtain the ground state energy and
the polaron contribution to the inverse effective mass for the 2TO polaron.
The ground state energy $\Delta E^{\left(  2\mathrm{TO}\right)  }\left(
P\right)  $ is plotted in Fig. \ref{fig:gsenergy}. Results of the present
method are labeled \textquotedblleft SPS\textquotedblright\ which stands for
the \textquotedblleft squeezed phonon state\textquotedblright\ approach, since
that is indeed what we rely on through the Bogoliubov-Tyablikov diagonalization.

The polaronic energy shift resulting from the 2TO interaction depends not only
on the magnitude of the interaction but also on its sign. For $\alpha_{T}>0$,
its behavior resembles the repulsive polaron in atomic quantum gases
\cite{Kohstall}. The dependence of the self-energy on $\alpha_{T}$ is not
fully antisymmetric when changing the sign of $\alpha_{T}$, because both even
and odd terms contribute to the total energy.

The dashed lines show the first-order perturbation result for the ground state
energy determined by the averaging of the electron-phonon interaction term
with the Hamiltonian of free electrons and phonons,%
\begin{equation}
E_{weak}^{\left(  2\mathrm{TO}\right)  }=\left\langle H_{2\mathrm{TO}%
}\right\rangle _{0}=\frac{3}{8}\frac{\hbar^{2}k_{0}^{2}}{m}\alpha_{T}.
\label{WC}%
\end{equation}
The ground state energy determined from (\ref{DE2TO}) at $P=0$ analytically
tends to $\left\langle H_{2\mathrm{TO}}\right\rangle _{0}$ in the limit of
small coupling constant $\alpha_{T}$.%

\begin{figure}[ptbh]%
\centering
\includegraphics[
height=3.2872in,
width=3.9323in
]%
{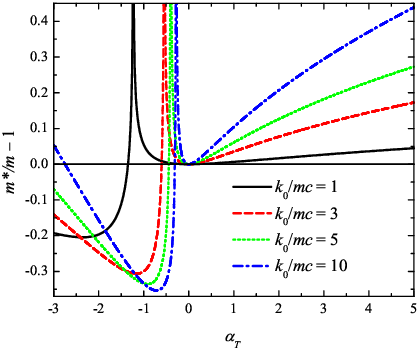}%
\caption{TO2-polaron effective mass as a function of the coupling constant
$\alpha_{T}$, calculated using different values of the the momentum cutoff
$k_{0}$.}%
\label{pmassPM}%
\end{figure}

The dependence of the effective mass on the coupling constant $\alpha$ is
shown in Fig. \ref{pmassPM}. For a positive coupling constant, the $\alpha
_{T}$ dependence of the effective mass is smooth and does not manifest any
specific feature. The 2TO polaron effective mass monotonically rises with an
increasing $\alpha_{T}$, as well as with an increasing phonon cutoff momentum
$k_{0}$.

The behavior of the effective mass at negative $\alpha_{T}$ is more
interesting. It exhibits a resonant divergent peak of the effective mass at
some value of $\alpha_{T}$. A divergence of the effective polaron mass at some
negative coupling strength of the quadratic interaction was also obtained in
our preceding work on a polaron in a finite-width band \cite{PRBL2023} using
the numerically exact Diagrammatic Monte Carlo method. This result has been
verified analytically in the case of an infinitely narrow conduction band,
which is known as the \emph{atomic limit}, also discussed in the next section.
The atomic limit results in an explicit analytic change of the phonon
frequency. When the coupling constant passes through a certain critical
negative value, this renormalized squared phonon frequency becomes negative,
which indicates a negative stiffness and an instability of the crystal lattice
possibly resulting in a structural phase transition. At the critical coupling
strength, the polaron effective mass diverges. Consequently, this also means
an instability of the polaron state.

When the width of the conduction band is not equal to zero, the effect of the
quadratic coupling is more complicated and can hardly be reduced to a
renormalization of the phonon frequency. Nevertheless, it is physically clear
that a quadratic polaron with any band width should become unstable when the
coupling strength reaches a sufficiently large negative value. This must be
also true for the polaron model with a parabolic electron dispersion treated
in this section. Therefore, the divergence of the polaron effective mass shown
in Fig. \ref{pmassPM} is attributed to the polaron instability. The present
approximation of squeezed phonon states gives formal solutions for coupling
strengths both above and below this critical value (denoted as $\alpha
_{T}^{\left(  c\right)  }$). However, we should only consider the range
$\alpha_{T}>\alpha_{T}^{\left(  c\right)  }$ as physically reasonable.

It should be noted that an account of a whole series of anharmonic terms in
the phonon Hamiltonian (see also discussions in Refs. \cite{Adolphs,Han2024})
might avoid this polaron instability, which may appear as an artifact of
restricting the phonon Hamiltonian to the quadratic order. However, even in
this case the instability would be an artifact of the quadratic polaron model
but not an artifact of the approximation of squeezed phonon states, because,
as shown in the next section, the DiagMC calculation and the approximation of
squeezed phonon states predict the same divergence of the polaron effective mass.

As was proven by Gerlach and L\"{o}wen \cite{Gerlach} who considered rigorous
relations for a polaron, \textquotedblleft phase transitions\textquotedblright%
\ when varying the coupling strength are forbidden for a rather wide class of
polarons, which however does not include all existing polaron models. Gerlach
and L\"{o}wen considered polarons with a linear electron-phonon coupling,
where the coupling amplitude depends only on the phonon momentum. The theorem
can be inapplicable even for a nonlocal linear electron-phonon interaction,
for example, the Peierls/Su-Schrieffer-Heeger polaron \cite{Prokofiev2021} or
for a polaron exposed to a short-range potential \cite{Gerlach}. Furthermore,
polarons with nonlinear electron-phonon interactions were beyond the scope of
the study in Ref. \cite{Gerlach}.

The polaron with a quadratic electron-phonon coupling of Ref. \cite{PRBL2023}
represents an example of an exactly solvable polaron model in the atomic
limit. It shows that the Gerlach-L\"{o}wen theorem can be inapplicable in the
case of a quadratic interaction.

Remarkably, there are kinks (discontinuities of the first derivative) in the
curve for the ground state energy in Fig. \ref{fig:gsenergy} at the same
critical values of the coupling constant where the 2TO polaron effective mass
diverges. As discussed above and in Sec. \ref{Sec:Scalar}, the physical origin
of these features consists in the polaron instability which appears when the
coupling strength of the quadratic interaction reaches some critical negative
value $\alpha_{T}^{\left(  c\right)  }$. In the parameter range for
$\alpha>\alpha_{T}^{\left(  c\right)  }$, the result of the approximation of
squeezed phonon states is well consistent with DiagMC simulations
(Fig.~\ref{fig:gsenergy} {(\emph{b})}) of a subset of the diagrammatic
expansion of the polaron Green function. The terms involved are all crossing
diagrams containing 2-loops and 3-loops, which are the dominant contribution
at small coupling. The inclusion of 3-loops results in a behavior closer to
the squeezed phonon state result compared to 2-loops alone. The implementation
of the DiagMC method is described in Appendix B.%

\begin{figure}[ptbh]%
\centering
\includegraphics[
height=3.3088in,
width=4.0136in
]%
{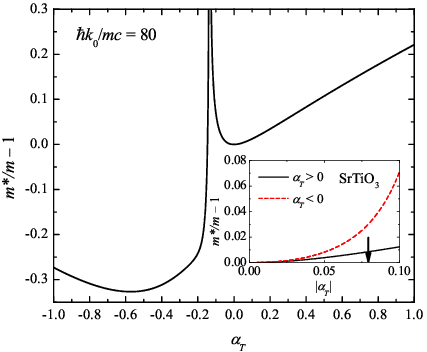}%
\caption{TO2-polaron effective mass as a function of the coupling constant
$\alpha_{T}$, calculated using the momentum cutoff $\hbar k_{0}=80mc$, which
approximately corresponds to the Brillouin zone edge in SrTiO$_{3}$.
\emph{Inset}: The effective mass in the weak-coupling range of $\alpha_{T}$.
The arrow indicates the value $\alpha_{T}\approx0.079$ obtained using the
parameters from Ref. \cite{Maslov}.}%
\label{fig:massSTO}%
\end{figure}

In Fig. \ref{fig:massSTO}, the 2TO polaron effective mass is shown as a
function of $\alpha_{T}$ choosing other parameters the same as in Ref.
\cite{Maslov} in order to see whether the 2TO interaction may be relevant for
polarons in strontium titanate. Kumar \emph{et al}. use $g_{2}$ as a fitting
parameter, and apply for numerics $g_{2}=0.92a_{0}^{3}$ where $a_{0}$ is the
lattice constant taken in Ref. \cite{Maslov} to be $a_{0}=3.9%
\operatorname{\text{\AA}}%
$. With other parameters from the same work, $c=6.6\times10^{5}%
\operatorname{cm}%
\operatorname{s}%
^{-1}$, the bare electron band mass $m=1.8m_{0}$ (where $m_{0}$ is the
electron mass in vacuum), and with $\hbar\omega_{\text{LO}}=0.0987%
\operatorname{eV}%
$ \cite{STO2010}, we estimate the dimensionless 2TO coupling constant in
SrTiO$_{3}$ as $\alpha_{T}\approx0.079$. This value is indicated by the arrow
in the inset of Fig. \ref{fig:massSTO}. The phonon cutoff momentum is chosen
here as the edge of the Brillouin zone, $k_{0}=\pi/a_{0}$, which gives us the
dimensionless cutoff value $p_{0}=\hbar k_{0}/mc\approx80$.

As can be seen from Fig. \ref{fig:massSTO}, the relative contribution of the
2TO interaction to the polaron mass in SrTiO$_{3}$ is relatively small with
respect to the Fr\"{o}hlich polaron mass, which can be estimated as
$m_{F}^{\ast}/m-1\approx\alpha/6\approx0.35$ \cite{STO2010}. However the 2TO
contribution is not negligible. Moreover, if it may appear that the coupling
constant in SrTiO$_{3}$ is negative, the value $\left\vert \alpha
_{T}\right\vert \approx0.079$ lies rather close to the resonance obtained in
the present calculation. This may explain larger values for the effective mass
of a \textquotedblleft dressed\textquotedblright\ electron obtained in
spectroscopic measurements with respect to that which follows from our
calculations (predicting $\alpha\approx2.1$).

Figures \ref{pmassPM} and \ref{fig:massSTO} show that the critical negative
coupling $\alpha_{T}^{\left(  c\right)  }$ gradually decreases in magnitude
when increasing the momentum cutoff $k_{0}$. If the cutoff value tends to
infinity, it means physically an infinite increase of a negative contribution
to the renormalized squared phonon frequency at any finite $\alpha_{T}$. Thus
we can suggest that in the limit $k_{0}\rightarrow\infty$, the critical
coupling constant $\alpha_{T}^{\left(  c\right)  }$ must tend to zero.

\section{Including nonparabolicity and finite width of the conduction band
\label{Sec:Scalar}}

\subsection{Model}

The most important point of this subsection consists in a generalization of
the method using displaced squeezed phonon states to the polaron in a
nonparabolic finite-width conduction band. The generalized method satisfies
periodic boundary conditions in the Brillouin zone and consequently it is not
restricted to a small polaron momentum. Thus, it allows for a description of
both large and small polarons.

The treatment in the present work is performed for the polaron model
introduced in Ref. \cite{PRBL2023}. The polaron problem is treated using the
Hamiltonian in the lattice representation for a simple cubic lattice, written
as $H=H_{0}+H_{e-ph}$ where
\begin{align}
H_{0}  &  =-t\sum_{\left\langle \mathbf{i}^{\prime},\mathbf{i}\right\rangle
}\sum_{\sigma=\pm1/2}c_{\mathbf{i}^{\prime},\sigma}^{\dag}c_{\mathbf{i}%
,\sigma}+\sum_{\mathbf{i}^{\prime},\mathbf{i}}\omega_{0}\left(  b_{\mathbf{i}%
^{\prime}}^{\dag}b_{\mathbf{i}}+\frac{1}{2}\right)  ,\\
H_{e-ph}  &  =\frac{\omega_{0}}{4}g\sum_{\mathbf{i}}n_{\mathbf{i}%
}B_{\mathbf{i}}^{2}, \label{Eph}%
\end{align}
with%
\begin{align}
B_{\mathbf{i}}  &  =b_{\mathbf{i}}+b_{\mathbf{i}}^{\dag},\\
n_{\mathbf{i}}  &  =\sum_{\sigma=\pm1/2}c_{\mathbf{i},\sigma}^{\dag
}c_{\mathbf{i},\sigma}.
\end{align}
In reciprocal space the lattice corresponds to a discrete finite set of wave
vectors,
\begin{equation}
-\frac{\pi}{a_{0}}\leq\left(  k_{j},q_{j}\right)  <\frac{\pi}{a_{0}}%
,\quad\left(  k_{j},q_{j}\right)  =\frac{2\pi i_{j}}{L}\quad\left(
j=x,y,z\right)  , \label{qj}%
\end{equation}
where $L=N^{1/3}a_{0}$ is the size of the system ($V=L^{3}=Na_{0}^{3}$),
$a_{0}$ being the lattice constant, and $i_{j}$ are integers.

In the momentum representation, applying the discrete Fourier transform%
\begin{align}
c_{\mathbf{i},\sigma}  &  =\frac{1}{N^{1/2}}\sum_{\mathbf{k}}c_{\mathbf{k}%
,\sigma}e^{i\mathbf{k}\cdot\mathbf{r}_{\mathbf{i}}},\quad c_{\mathbf{i}%
,\sigma}^{\dag}=\frac{1}{N^{1/2}}\sum_{\mathbf{k}}c_{\mathbf{k},\sigma}^{\dag
}e^{-i\mathbf{k}\cdot\mathbf{r}_{\mathbf{i}}},\\
B_{\mathbf{q}}  &  =\frac{1}{N^{1/2}}\sum_{\mathbf{i}}e^{-i\mathbf{q}%
\cdot\mathbf{r}_{\mathbf{i}}}B_{\mathbf{i}},\qquad B_{\mathbf{i}}=\frac
{1}{N^{1/2}}\sum_{\mathbf{q}}e^{i\mathbf{q}\cdot\mathbf{r}_{\mathbf{i}}%
}B_{\mathbf{q}},
\end{align}
the Hamiltonian $H_{0}$ takes the form
\begin{equation}
H_{0}=\sum_{\mathbf{k}}\varepsilon\left(  \mathbf{k}\right)  \sum_{\sigma
}c_{\mathbf{k},\sigma}^{\dag}c_{\mathbf{k},\sigma}+\sum_{\mathbf{q}}\omega
_{0}\left(  b_{\mathbf{q}}^{\dag}b_{\mathbf{q}}+\frac{1}{2}\right)
\label{H0b}%
\end{equation}
with $\varepsilon\left(  \mathbf{k}\right)  $ (counted from the middle of the
conduction band) given by%
\begin{equation}
\varepsilon\left(  \mathbf{k}\right)  =-2t\sum_{j=x,y,z}\cos(k_{j}a_{0}).
\label{e2a}%
\end{equation}
The quadratic interaction Hamiltonian becomes%
\begin{equation}
H_{e-ph}=\frac{\omega_{0}}{4}g\frac{1}{N}\sum_{\mathbf{q}}\sum_{\mathbf{q}%
^{\prime}}\sum_{\mathbf{k}}\sum_{\sigma}c_{\mathbf{k}+\mathbf{q}%
-\mathbf{q}^{\prime},\sigma}^{\dag}c_{\mathbf{k},\sigma}B_{\mathbf{q}%
}B_{\mathbf{q}^{\prime}}^{\dag} \label{Heph}%
\end{equation}
with $B_{\mathbf{q}}=b_{\mathbf{q}}+b_{-\mathbf{q}}^{\dag}$.

Expanding the band dispersion up to quadratic order, $\varepsilon\left(
\mathbf{k}\right)  =a_{0}^{2}t~k^{2}+O\left(  k^{4}\right)  $, yields the
relation between the bandwidth parameter $t$ and the band mass $m_{b}$:%
\begin{equation}
t=\frac{\hbar^{2}}{2m_{b}a_{0}^{2}},\qquad m_{b}=\frac{\hbar^{2}}{2a_{0}^{2}t}%
\end{equation}
The band width for the tight-binding model is $W\equiv\max\left[
\varepsilon\left(  \mathbf{k}\right)  \right]  -\min\left[  \varepsilon\left(
\mathbf{k}\right)  \right]  =12t$. Further on, we set $\hbar=1$. The other
units will be set below.

For a single polaron, the parts of the single-polaron Hamiltonian are
\begin{align}
H_{0}\left(  \mathbf{p},\left\{  b_{\mathbf{q}}^{\dag},b_{\mathbf{q}}\right\}
\right)   &  =\varepsilon\left(  \mathbf{p}\right)  +\sum_{\mathbf{q}}%
\omega_{0}\left(  b_{\mathbf{q}}^{\dag}b_{\mathbf{q}}+\frac{1}{2}\right)
,\label{H0a}\\
H_{e-ph}\left(  \mathbf{r},\left\{  b_{\mathbf{q}}^{\dag},b_{\mathbf{q}%
}\right\}  \right)   &  =\frac{\omega_{0}}{4}g\frac{1}{N}\sum_{\mathbf{q}}%
\sum_{\mathbf{q}^{\prime}}e^{i\left(  \mathbf{q}-\mathbf{q}^{\prime}\right)
\cdot\mathbf{r}}B_{\mathbf{q}}B_{\mathbf{q}^{\prime}}^{\dag}. \label{Hepha}%
\end{align}

The first Lee-Low-Pines transformation, as in Sec. \ref{Sec:2TO}, leads to an
electron coordinate-free Hamiltonian. In order to apply the
Bogoliubov-Tyablikov transformation, this Hamiltonian is rewritten in terms of
real phonon coordinates,
\begin{equation}
\mathcal{H}=\varepsilon\left(  \mathbf{P}-\mathbf{Q}\right)  +\sum
_{\mathbf{q}}\omega_{0}\left(  b_{\mathbf{q}}^{\dag}b_{\mathbf{q}}+\frac{1}%
{2}\right)  +\frac{\omega_{0}g}{4N}\left(  \sum_{\mathbf{q}}\left(
b_{\mathbf{q}}+b_{\mathbf{q}}^{\dag}\right)  \right)  ^{2}, \label{HCF2}%
\end{equation}
where the total phonon momentum is $\mathbf{Q}=\sum_{\mathbf{q}}%
\mathbf{q}b_{\mathbf{q}}^{\dag}b_{\mathbf{q}}$. Representing $\varepsilon
\left(  \mathbf{P}-\mathbf{Q}\right)  $ through the normal products of phonon
second quantization operators, we arrive at the result:%
\begin{align}
\varepsilon\left(  \mathbf{P}-\mathbf{Q}\right)   &  =-t\sum_{j=1}^{3}\left[
e^{ia_{0}P_{j}}\mathtt{N}\exp\left(  \sum_{\mathbf{q}}\left(  e^{-ia_{0}q_{j}%
}-1\right)  b_{\mathbf{q}}^{\dag}b_{\mathbf{q}}\right)  \right. \nonumber\\
&  \left.  +e^{-ia_{0}P_{j}}\mathtt{N}\exp\left(  \sum_{\mathbf{q}}\left(
e^{ia_{0}q_{j}}-1\right)  b_{\mathbf{q}}^{\dag}b_{\mathbf{q}}\right)  \right]
. \label{EnerPQ}%
\end{align}
where $\mathtt{N}\left(  \ldots\right)  $ denotes the normal form of second
quantization operators. When truncating the Taylor series of (\ref{EnerPQ}) in
powers of normal products of phonon operators up to the quadratic order, this
gives us the expression%
\begin{equation}
\varepsilon^{\left(  quad\right)  }\left(  \mathbf{P}-\mathbf{Q}\right)
=\varepsilon\left(  \mathbf{P}\right)  +\sum_{\mathbf{q}}\left[
\varepsilon\left(  \mathbf{P}-\mathbf{q}\right)  -\varepsilon\left(
\mathbf{P}\right)  \right]  b_{\mathbf{q}}^{\dag}b_{\mathbf{q}}. \label{Equad}%
\end{equation}
Consequently, the Hamiltonian (\ref{HCF2}) can be subdivided in the two parts%
\begin{equation}
\mathcal{H}=H_{0}+H_{I} \label{subdiv}%
\end{equation}
where the Hamiltonian $H_{0}$ is a quadratic form of phonon operators,%
\begin{equation}
H_{0}=\varepsilon\left(  \mathbf{P}\right)  +\sum_{\mathbf{q}}\frac{\omega
_{0}}{2}+\sum_{\mathbf{q}}\Omega_{\mathbf{q}}b_{\mathbf{q}}^{\dag
}b_{\mathbf{q}}+\frac{\omega_{0}g}{4N}\left(  \sum_{\mathbf{q}}\left(
b_{\mathbf{q}}+b_{\mathbf{q}}^{\dag}\right)  \right)  ^{2}, \label{H0c}%
\end{equation}
with the renormalized phonon frequency $\Omega_{\mathbf{q}}$,%
\begin{equation}
\Omega_{\mathbf{q}}=\omega_{0}+\varepsilon\left(  \mathbf{P}-\mathbf{q}%
\right)  -\varepsilon\left(  \mathbf{P}\right)  , \label{Wq}%
\end{equation}
and $H_{I}$ is a series of all higher-order terms beyond the quadratic
expansion,
\begin{equation}
H_{I}=\varepsilon\left(  \mathbf{P}-\mathbf{Q}\right)  -\varepsilon^{\left(
quad\right)  }\left(  \mathbf{P}-\mathbf{Q}\right)  . \label{H0d}%
\end{equation}

As can be explicitly seen from (\ref{EnerPQ}), both the total Hamiltonian
$\mathcal{H}$ and the quadratic Hamiltonian $H_{0}$ have the correct periodic
translation symmetry, the same as for the initial exact electron-phonon
Hamiltonian. Namely, $\varepsilon\left(  \mathbf{P}\right)  $, $\varepsilon
\left(  \mathbf{P}-\mathbf{Q}\right)  $ and $\Omega_{\mathbf{q}}$ are
invariant with respect to the periodic translations $P_{j}\rightarrow
P_{j}+2\pi/a_{0}$ and/or $q_{j}\rightarrow q_{j}+2\pi/a_{0}$. Consequently,
the present scheme exactly accounts for the boundary conditions of the
Brillouin zone both for an electron and phonons. When including both linear
and quadratic electron-phonon interactions, the full expansion of the kinetic
energy in powers of phonon operators is performed in the same way without
difficulties. Therefore this expansion gives us the straightforward extension
of the method of squeezed phonon states to a polaron in a nonparabolic
finite-width band. As a particular case, this provides the equivalent scheme
of displaced squeezed phonon approach for a small polaron.

\subsection{Self-energy of a polaron with a quadratic interaction}

The shift of the self-energy provided by the Bogoliubov-Tyablikov
diagonalization is determined in the same way as in Sec. \ref{Sec:2TO} and
gives
\begin{equation}
\Delta E\left(  \mathbf{P}\right)  =-\frac{1}{8\pi i}\oint_{C}ds~\frac
{1}{\sqrt{s}}\ln\left(  1-\frac{\omega_{0}g}{N}\sum_{\mathbf{q}}\frac
{\Omega_{\mathbf{q}}}{s-\Omega_{\mathbf{q}}^{2}}\right)  . \label{dEP}%
\end{equation}
The summation over $\mathbf{q}$ is performed within the first Brillouin zone
over sites of the reciprocal lattice with the number of sites $N=\frac
{V}{a_{0}^{3}}=\left(  2l\right)  ^{3}$ (with $L\equiv2la_{0}$). Thus the
present treatment is in fact for the lattice polaron rather than for the
continuum polaron in bulk. The subsequent numeric check shows that the
relative difference between the energies for the continuum and lattice
polarons becomes negligibly small already at relatively small $l$. For
example, the relative difference of the ground state energies calculated with
$l=10$ and $l=20$ is about $2.4\times10^{-9}$. Consequently, the lattice
representation very well reproduces the properties of a polaron in bulk even
at relatively small number of sites.

To obtain the particular case of the atomic limit (AL), the limiting
transition $t\rightarrow0$ can be taken explicitly for the leading term of the
ground state energy. For the ground state energy in this limit we can consider
the Hamiltonian%
\begin{equation}
\lim_{t\rightarrow0}H_{0}=\sum_{\mathbf{q}}\omega_{0}\left(  b_{\mathbf{q}%
}^{\dag}b_{\mathbf{q}}+\frac{1}{2}\right)  +\frac{\omega_{0}g}{4N}\left(
\sum_{\mathbf{q}}\left(  b_{\mathbf{q}}+b_{\mathbf{q}}^{\dag}\right)  \right)
^{2}. \label{limH0}%
\end{equation}
For the numeric calculation, we apply the expression (\ref{dEP}), which is
simplified in the atomic limit to the analytic expression for the ground-state
energy, which is exact for (\ref{limH0}):%
\begin{align}
\Delta E_{0}^{\left(  AL\right)  }  &  =-\frac{\omega_{0}}{4\pi i}\oint%
_{C}dz\ln\left(  1-\frac{g}{z^{2}-1}\right) \nonumber\\
&  =\frac{1}{2}\omega_{0}\left(  \sqrt{g+1}\Theta\left(  g+1\right)
-1\right)  , \label{AL}%
\end{align}
where $\Theta\left(  g+1\right)  $ is the Heaviside step function.
Equivalently, the Hamiltonian (\ref{limH0}) can be exactly diagonalized by the
Bogoliubov-Tyablikov canonical transformation similarly to (\ref{H0c}). In the
atomic limit, the value $g=-1$ indicates the polaron instability as discussed
above and in Ref. \cite{PRBL2023}.%

\begin{figure}[ptbh]%
\centering
\includegraphics[
height=3in,
width=5.6749in
]%
{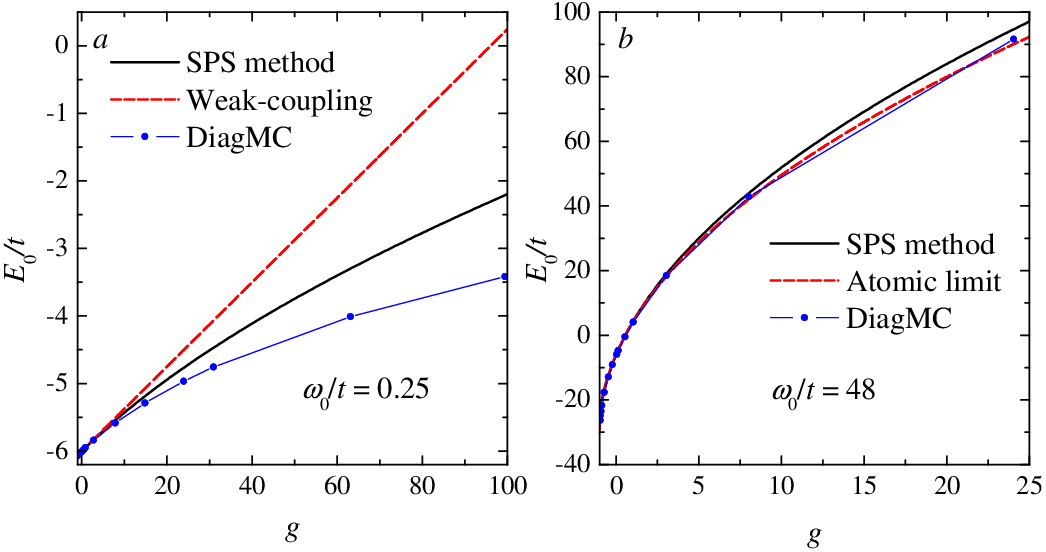}%
\caption{Ground state energy of the quadratic polaron in the adiabatic regime
with $\omega_{0}=0.25t$ (\emph{a}) and in the antiadiabatic regime with
$\omega_{0}=48t$ (\emph{b}) as a function of the strength $g$ of the quadratic
electron-phonon coupling calculated within the present approach (solid curve)
and by DiagMC (dots). The DiagMC data are from Ref. \cite{PRBL2023}. In panel
(\emph{a}), the dashed line shows the weak-coupling limit for the ground state
energy within the first-order perturbation theory. In panel (\emph{b}), the
dashed curve shows the ground-state energy in the atomic limit $t\rightarrow0$
calculated using the expression (\ref{AL}).}%
\label{fig:PolE0}%
\end{figure}

In Fig.~\ref{fig:PolE0} (\emph{a}), we plot the ground state polaron energy of
a polaron with a quadratic interaction as a function of the coupling constant
$g$ in the adiabatic regime, with $\omega_{0}=0.25t$. The obtained ground
state energy is compared with the DiagMC data of Ref.~\cite{PRBL2023} shown by
full dots.

As we can see from Fig.~\ref{fig:PolE0} (\emph{a}), the qualitative behavior
of the ground state energy within the extended squeezed phonon approach is
similar to that obtained using DiagMC calculations. In the adiabatic regime
the extension of the squeezed phonon method to the polaron in the
tight-binding conduction band provides polaron ground state energy values in
between the weak-coupling results and the DiagMC data, being closer to DiagMC
rather than to the weak-coupling result. For a weak and intermediate coupling
strength ($g\lesssim10$), the agreement between the squeezed phonon
approximation and DiagMC results seems to be rather good.

In Fig.~\ref{fig:PolE0} (\emph{b}), the ground state energy is calculated for
$\omega_{0}/t=48$, which corresponds to the antiadiabatic regime. In the
antiadiabatic regime, when $\omega_{0}\gg t$, the agreement between the
current method with squeezed phonon states and DiagMC for the ground state
energy appears to be better than in the adiabatic regime. This is explained by
the fact that in the limit $t\rightarrow0$, the coordinate-free Hamiltonian
(\ref{HCF2}) tends to a quadratic form which is \emph{exactly} diagonalized by
the Bogoliubov-Tyablikov transformation. For the comparison, the result of
this limiting transition is shown by the red dashed curve in Fig.
\ref{fig:PolE0} (\emph{b}). The ground state energy calculated using the
approximation of squeezed phonon states consistently registers higher values
compared to the results derived from the full DiagMC series, as distinct from
the partial DiagMC summation in Sec. \ref{Sec:2TO}. This trend is not observed
in the atomic limit in comparison with DiagMC results due to the finite width
of the conduction band. Under strong coupling conditions, the ground state
energy within the atomic limit may fall below the DiagMC results for $t\neq0$,
as depicted in Fig. \ref{fig:PolE0} (b).%

\begin{figure}[ptbh]%
\centering
\includegraphics[
height=2.7908in,
width=5.278in
]%
{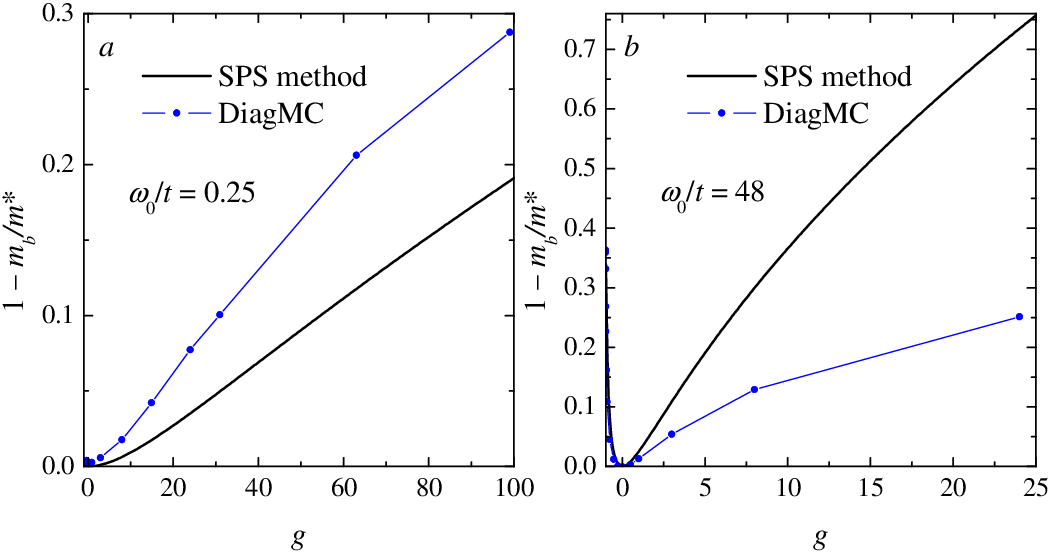}%
\caption{Parameter $\kappa=1-m_{b}/m^{\ast}$ of the quadratic polaron in the
adiabatic regime with $\omega_{0}=0.25t$ (\emph{a}) and in the antiadiabatic
regime with $\omega_{0}=48t$ (\emph{b}) as a function of the coupling strength
$g$ calculated using the SPS method (solid curves) and by DiagMC, Ref.
\cite{PRBL2023} (dots).}%
\label{fig:Pmass}%
\end{figure}

In Fig.~\ref{fig:Pmass}, we plot the parameter $\kappa=1-m_{b}/m^{\ast}$ which
determines the polaron effective mass $m^{\ast}$, in the adiabatic (\emph{a})
and antiadiabatic (\emph{b}) regimes. The parameter $\kappa$ is the
coefficient at $k^{2}$ in the series expansion of the momentum-dependent
polaron energy $E_{p}\left(  \mathbf{k}\right)  $ in powers of $\mathbf{k}$.
For the simple cubic tight-binding band used in the present work and in
Ref.~\cite{PRBL2023}, the electron and polaron effective masses are isotropic.

The dependence of $\kappa$ as a function of the coupling constant $g$ is
qualitatively similar to that extracted from the DiagMC data. However,
quantitatively there is a difference between our results and
Ref.~\cite{PRBL2023}. In the adiabatic regime, the present calculation
underestimates the polaron mass with respect to the DiagMC result. On the
contrary, in the antiadiabatic regime we can see an overestimation of $\kappa$
given by the present method with respect to DiagMC.%

\begin{figure}[ptbh]%
\centering
\includegraphics[
height=4.9476in,
width=4.4806in
]%
{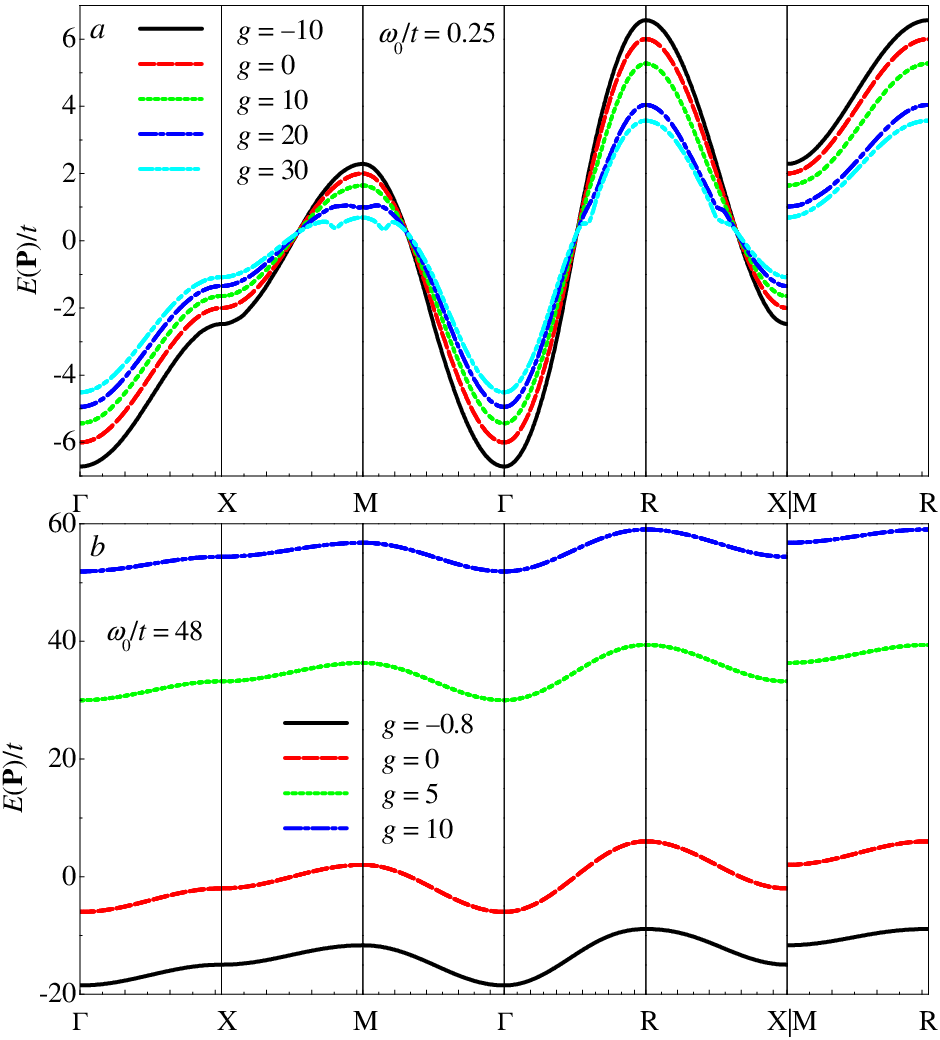}%
\caption{Band dispersion of the polaron with a quadratic interaction in the
adiabatic case with $\omega_{0}/t=0.25$ (\emph{a}) and in the antiadiabatic
case with $\omega_{0}/t=48$ (\emph{b}) along the path for the cubic lattice
$\Gamma-X-M-\Gamma-R-X|M-R$ for different values of the coupling strength.}%
\label{fig:DispPolaron}%
\end{figure}

Because the approximated coordinate-free Hamiltonian keeps the translation
symmetry of the initial electron-phonon Hamiltonian, it allows us to calculate
the polaron energy $E_{p}\left(  \mathbf{k}\right)  $ in the whole Brillouin
zone. In Fig.~\ref{fig:DispPolaron} (\emph{a}), the band dispersion is shown
for the polaron with a quadratic interaction for several values of the
coupling constant $g$ (including $g=0$ which corresponds to the bare band
electron) along the standard path for the cubic lattice: $\Gamma
-X-M-\Gamma-R-X|M-R$ in the adiabatic regime with $\omega_{0}/t=0.25$. As can
be seen from the figure, the quadratic electron-phonon interaction leads to a
narrowing of the conduction band with respect to that of the bare electron.
For sufficiently high coupling strengths, the polaron self-energy exhibits
non-monotonic behavior at large momentum close to the point $M$, so that local
minima appear along the chosen path. Consequently, the polaron band dispersion
is more complicated than the electron band dispersion.

The polaron band dispersion in the antiadiabatic regime shown in
Fig.~\ref{fig:DispPolaron} (\emph{b}) looks differently from the
momentum-dependent energy in the adiabatic regime. First, here the polaron
shift of the energy is relatively large with respect to the electron
bandwidth. As a result, the polaron effect on the energy in the antiadiabatic
regime is expressed through the shift of the whole band rather than through a
renormalization of the bandwidth. Thus in the antiadiabatic regime the top and
the bottom of the conduction band shift in the same directions, while in the
adiabatic case they shift in the opposite directions.

Looking back at Figs \ref{fig:PolE0}(\emph{b}) and \ref{fig:Pmass}(\emph{b}),
we can note that the method of squeezed phonon states is highly effective at
approximating the ground-state energy in the deep anti-adiabatic regime, but
it performs rather poorly in quantitatively estimating the effective mass.
Observing the polaron dispersion in Fig. \ref{fig:DispPolaron}, we can
conclude that in the deep antiadiabatic regime the intraband dispersion of the
polaron self-energy (as well the conduction band width) is small with respect
to the shift of the whole band. The good agreement between the method of
squeezed phonon states and DiagMC on the ground-state energy in the deep
antiadiabatic regime is explained by the fact that the expression
(\ref{limH0}) represents the atomic limit not only for the truncated
Hamiltonian (\ref{H0c}) but also for the exact one (\ref{HCF2}). As the
variation of the polaron self-energy within the conduction band in the deep
antiadiabatic regime constitutes only a small part of the total self-energy,
the relative error for the effective mass in this regime can be substantially
larger than for the ground-state energy. Despite this quantitative deviation
from the DiagMC results, the method of squeezed phonon states rather well
captures the behavior of the effective mass at $g<0$ and the value of the
critical negative coupling strength indicating the polaron instability.

\section{Conclusions \label{sec:Conclusions}}

One of the primary outcomes of this treatment involves incorporating quadratic
electron-phonon interaction into the squeezed phonon states scheme, which
appears straightforward. The resulting dependence of the ground state energy
and the effective mass of a polaron, arising from the quadratic interaction
between an electron and TO-phonons, exhibits notable differences for positive
and negative coupling constants. In the regime of positive coupling strengths,
we observe behavior typical of a repulsive polaron, without irregularities.
However, for a negative coupling constant, the polaron's effective mass
diverges at a critical value of the negative coupling strength. This
divergence indicates a polaron instability, akin to what has been described in
Ref.~\cite{PRBL2023}.

An advantageous feature of the current modification to the squeezed phonon
states method lies in its ability to calculate the polaron band dispersion
across the entire Brillouin zone, extending beyond the assumption of small
polaron momentum and the quadratic expansion in powers of polaron momentum,
and keeping the periodic boundary conditions exactly. This represents the
second key result of our present work.

Even for a parabolic band, achieving a comprehensive variational treatment is
inherently complex \cite{Shchadilova}. The feasibility of obtaining a
tractable form for the complete correlated Gaussian wavefunction remains
uncertain in the context of a nonparabolic conduction band. This unresolved
question serves as the focus of subsequent studies.

The analytic method employed in this work enables the investigation of various
polaron characteristics across a broad range of parameters, including coupling
strength. Notably, it complements the numeric DiagMC treatment. Our method's
predictions align qualitatively with DiagMC results, highlighting features
such as the instability of the polaron with quadratic interaction at specific
negative coupling strengths.

\begin{acknowledgments}
This work is supported financially by the Research Foundation--Flanders,
Projects No. GOH1122N, No. G061820N, and No. G060820N, and by the University
Research Fund (BOF) of the University of Antwerp. M.H. acknowledges funding by
the Research Foundation--Flanders via a postdoctoral fellowship (Grant No.
1224724N). Financial support from the Austrian Science Fund (FWF) Project No.
I 4506 (FWO- FWF joint project) is gratefully acknowledged. A.S.M.
acknowledges the support of the FrustKor project financed by the European
Union through the National Recovery and Resilience Plan 2021-2026 (NRPP) and
JSPS KAKENHI Grant Kiban A:24H00197.
\end{acknowledgments}

\appendix

\section{Derivation of the polaron self-energy}

The off-diagonal component of the quadratic Hamiltonian (\ref{H0}) and
(\ref{H0c}) remains irrelevant to the energy within the Lee-Low-Pines
approximation. The Bogoliubov-Tyablikov canonical transformation, as described
in \cite{Gross,Tulub1962,Tulub2015}, is the third transformation,%
\begin{align}
b_{\mathbf{q}}^{\left(  a\right)  }  &  =\frac{1}{\sqrt{V}}\sum_{\mathbf{q}%
^{\prime},a^{\prime}}\left(  u_{\mathbf{qq}^{\prime}}^{\left(  aa^{\prime
}\right)  }\beta_{\mathbf{q}^{\prime}}^{\left(  a^{\prime}\right)
}+v_{\mathbf{qq}^{\prime}}^{\left(  aa^{\prime}\right)  \ast}\beta
_{\mathbf{q}^{\prime}}^{\left(  a^{\prime}\right)  \dag}\right)  ,\nonumber\\
b_{\mathbf{q}}^{\left(  a\right)  \dag}  &  =\frac{1}{\sqrt{V}}\sum
_{\mathbf{q}^{\prime},a^{\prime}}\left(  u_{\mathbf{qq}^{\prime}}^{\left(
aa^{\prime}\right)  \ast}\beta_{\mathbf{q}^{\prime}}^{\left(  a^{\prime
}\right)  \dag}+v_{\mathbf{qq}^{\prime}}^{\left(  aa^{\prime}\right)  }%
\beta_{\mathbf{q}^{\prime}}^{\left(  a^{\prime}\right)  }\right)  . \label{BT}%
\end{align}
This mixes the creation and annihilation phonon operators, and corresponds to
a transformation from the original phonon states to squeezed phonon states.
This is why we refer to this approach as the \textquotedblleft squeezed phonon
state\textquotedblright\ approach. When the Bogoliubov-Tyablikov
transformation is used in conjunction with the displacement operator $S_{2}$,
this leads to displaced squeezed phonon states, which are useful when both a
linear-harmonic interaction and a quadratic interaction are present. The
matrix elements of the Bogoliubov-Tyablikov unitary transformation are chosen
in order to diagonalize the Hamiltonian $H_{0}$. After applying (\ref{BT}) to
(\ref{H0c}), the resulting Hamiltonian can be written as:%
\begin{equation}
H_{0}=E_{0}+\sum_{a,\mathbf{q}}\hbar\nu_{\mathbf{q}}^{\left(  a\right)  }%
\beta_{\mathbf{q}}^{\left(  a\right)  \dag}\beta_{\mathbf{q}}^{\left(
a\right)  }+\Delta E,
\end{equation}
where $\nu_{\mathbf{q}}^{\left(  a\right)  }$ are eigenfrequencies, and
$\Delta E$ is the polaron energy shift,%
\begin{equation}
\Delta E=\frac{1}{2}\sum_{a,\mathbf{q}}\hbar\left(  \nu_{\mathbf{q}}^{\left(
a\right)  }-\Omega_{\mathbf{q}}^{\left(  a\right)  }\right)  .
\end{equation}

In order to obtain the polaron self-energy within the squeezed phonon states
method for $H_{0}$, we do not need an explicit form of eigenfrequencies and
matrix elements. The self-energy can be derived using a scheme described by
Wentzel \cite{Wentzel}. First, the Hamiltonian $H_{0}$ is rewritten in terms
of phonon coordinates and momenta,
\begin{align}
Q_{\mathbf{q}}^{\left(  a\right)  }  &  =\sqrt{\frac{\hbar}{2m\Omega
_{\mathbf{q}}^{\left(  a\right)  }}}\left(  b_{\mathbf{q}}^{\left(  a\right)
\dag}+b_{\mathbf{q}}^{\left(  a\right)  }\right) \nonumber\\
P_{\mathbf{q}}^{\left(  a\right)  }  &  =-i\sqrt{\frac{\hbar m\Omega
_{\mathbf{q}}^{\left(  a\right)  }}{2}}\left(  b_{\mathbf{q}}^{\left(
a\right)  }-b_{\mathbf{q}}^{\left(  a\right)  \dag}\right)  \label{coords}%
\end{align}
leading to%
\begin{align}
H_{0}  &  =E_{0}+\sum_{a,\mathbf{q}}\left(  \frac{\left(  P_{\mathbf{q}%
}^{\left(  a\right)  }\right)  ^{2}}{2m}+\frac{m\left(  \Omega_{\mathbf{q}%
}^{\left(  a\right)  }\right)  ^{2}}{2}\left(  Q_{\mathbf{q}}^{\left(
a\right)  }\right)  ^{2}\right) \nonumber\\
&  +\hbar\mathbf{W}_{1}^{2}+\frac{g_{2}m}{\hbar}\mathbf{W}_{2}^{2}-\frac{1}%
{2}\sum_{a,\mathbf{q}}\hbar\Omega_{\mathbf{q}}^{\left(  a\right)  },
\label{H0a1}%
\end{align}
with the collective coordinates%
\begin{equation}
\mathbf{W}_{1}=\sum_{a,\mathbf{q}}\mathbf{q}\sqrt{\Omega_{\mathbf{q}}^{\left(
a\right)  }}f_{\mathbf{q}}^{\left(  a\right)  }Q_{\mathbf{q}}^{\left(
a\right)  },\;\mathbf{W}_{2}=\sum_{\mathbf{q},a=1,2}\frac{\mathbf{e}%
_{\mathbf{q}}^{\left(  a\right)  }\varkappa_{\mathbf{q}}\sqrt{\Omega
_{\mathbf{q}}^{\left(  a\right)  }}}{\sqrt{V}}Q_{\mathbf{q}}^{\left(
a\right)  }. \label{coll}%
\end{equation}
The equation for eigenfrequencies and eigenvectors of the quadratic form
(\ref{H0a1}) is determined in the standard way:%
\begin{equation}
\sum_{a^{\prime},\mathbf{q}^{\prime}}M_{\mathbf{q}^{\prime},\mathbf{q}%
}^{\left(  a,a^{\prime}\right)  }\left(  \omega\right)  Q_{\mathbf{q}^{\prime
}}^{\left(  a^{\prime}\right)  }\left(  \omega\right)  =0, \label{Eigen}%
\end{equation}
where the elements of the matrix $\mathbb{M}\left(  \omega\right)  =\left\Vert
M_{\mathbf{q}^{\prime},\mathbf{q}}^{\left(  a,a^{\prime}\right)  }\left(
\omega\right)  \right\Vert $ are
\begin{align}
M_{\mathbf{q}^{\prime},\mathbf{q}}^{\left(  a,a^{\prime}\right)  }\left(
\omega\right)   &  =\delta_{a^{\prime},a}\delta_{\mathbf{q}^{\prime
},\mathbf{q}}\left[  \omega^{2}-\left(  \Omega_{\mathbf{q}}^{\left(  a\right)
}\right)  ^{2}\right]  -2\hbar\left(  \mathbf{q}\otimes\mathbf{q}^{\prime
}\right)  \sqrt{\Omega_{\mathbf{q}}^{\left(  a\right)  }\Omega_{\mathbf{q}%
^{\prime}}^{\left(  a^{\prime}\right)  }}f_{\mathbf{q}}^{\left(  a\right)
}f_{\mathbf{q}^{\prime}}^{\left(  a^{\prime}\right)  }\nonumber\\
&  -\left(  1-\delta_{a,3}\right)  \left(  1-\delta_{a^{\prime},3}\right)
\frac{2g_{2}}{\hbar V}\left(  \mathbf{e}_{\mathbf{q}}^{\left(  a\right)
}\otimes\mathbf{e}_{\mathbf{q}^{\prime}}^{\left(  a^{\prime}\right)  }\right)
\varkappa_{\mathbf{q}}\varkappa_{\mathbf{q}^{\prime}}\sqrt{\Omega_{\mathbf{q}%
}^{\left(  a\right)  }\Omega_{\mathbf{q}^{\prime}}^{\left(  a^{\prime}\right)
}}. \label{Matr}%
\end{align}
The eigenfrequencies are the roots of the equation
\begin{equation}
\det\mathbb{M}\left(  \omega\right)  =0. \label{detMeq0}%
\end{equation}
A reduced set of equations for collective coordinates (\ref{coll}) can be
extracted from the full set of equations (\ref{Eigen}) when we divide the
equation by $\left[  \omega^{2}-\left(  \Omega_{\mathbf{q}}^{\left(  a\right)
}\right)  ^{2}\right]  $ and perform summations over $\mathbf{q}$ with
different weight coefficients. It results in the matrix equation
\begin{equation}
\mathbb{A}\left(  \omega\right)  \mathbf{W}\left(  \omega\right)  =0,
\label{Red}%
\end{equation}
where $\mathbf{W}\left(  \omega\right)  $ is a 6-dimensional vector, which is
given in a block form by:%
\[
\mathbf{W}\left(  \omega\right)  =\left(
\begin{array}
[c]{c}%
\mathbf{W}_{1}\left(  \omega\right) \\
\mathbf{W}_{2}\left(  \omega\right)
\end{array}
\right)  .
\]
The matrix $\mathbb{A}\left(  \omega\right)  $ can be written as the block
matrix%
\begin{equation}
\mathbb{A}\left(  \omega\right)  =\left(
\begin{array}
[c]{cc}%
\mathbb{A}^{\left(  {\text{LO}}\right)  }\left(  \omega\right)  &
\mathbb{A}^{\left(  mix\right)  }\left(  \omega\right) \\
\left[  \mathbb{A}^{\left(  mix\right)  }\left(  \omega\right)  \right]  ^{T}
& \mathbb{A}^{\left(  2\mathrm{TO}\right)  }\left(  \omega\right)
\end{array}
\right)  , \label{Am}%
\end{equation}
with the matrices%
\begin{align}
\mathbb{A}^{\left(  {\text{LO}}\right)  }\left(  \omega\right)   &
=\mathbb{I}-2\hbar\sum_{\mathbf{q},a=1,2,3}\alpha_{\mathbf{q}}^{\left(
a\right)  }\left(  \mathbf{q}\otimes\mathbf{q}\right)  ,\label{ALO}\\
\mathbb{A}^{\left(  2\mathrm{TO}\right)  }\left(  \omega\right)   &
=\mathbb{I}-\frac{2g_{2}}{\hbar V}\sum_{\mathbf{q},a=1,2}\lambda_{\mathbf{q}%
}^{\left(  a\right)  }\left(  \mathbf{e}_{\mathbf{q}}^{\left(  a\right)
}\otimes\mathbf{e}_{\mathbf{q}}^{\left(  a\right)  }\right)  ,\label{ATO}\\
\mathbb{A}^{\left(  mix\right)  }\left(  \omega\right)   &  =-\frac{2g_{2}%
}{\hbar\sqrt{V}}\sum_{\mathbf{q},a=1,2}\gamma_{\mathbf{q}}^{\left(  a\right)
}\left(  \mathbf{q}\otimes\mathbf{e}_{\mathbf{q}}^{\left(  a\right)  }\right)
, \label{Amix}%
\end{align}
and weight functions
\begin{equation}
\alpha_{\mathbf{q}}^{\left(  a\right)  }\left(  \omega\right)  =\frac
{\Omega_{\mathbf{q}}^{\left(  a\right)  }\left(  f_{\mathbf{q}}^{\left(
a\right)  }\right)  ^{2}}{\omega^{2}-\left(  \Omega_{\mathbf{q}}^{\left(
a\right)  }\right)  ^{2}},\lambda_{\mathbf{q}}^{\left(  a\right)  }\left(
\omega\right)  =\frac{\varkappa_{\mathbf{q}}^{2}\Omega_{\mathbf{q}}^{\left(
a\right)  }}{\omega^{2}-\left(  \Omega_{\mathbf{q}}^{\left(  a\right)
}\right)  ^{2}},\gamma_{\mathbf{q}}^{\left(  a\right)  }\left(  \omega\right)
=\frac{\varkappa_{\mathbf{q}}\Omega_{\mathbf{q}}^{\left(  a\right)
}f_{\mathbf{q}}^{\left(  a\right)  }}{\omega^{2}-\left(  \Omega_{\mathbf{q}%
}^{\left(  a\right)  }\right)  ^{2}}. \label{coefs1}%
\end{equation}
The eigenvalues of the phonon energy are determined by the equation%
\begin{equation}
\det\mathbb{A}\left(  \omega\right)  =0.
\end{equation}

A mixing of the Fr\"{o}hlich and 2TO interactions is provided by non-diagonal
blocks of the matrix (\ref{Am}). For $P=0$, they are exactly equal to zero due
to symmetry. Therefore, for the ground-state energy the Fr\"{o}hlich and 2TO
contributions are completely decoupled within the approach of squeezed phonon
states. For a nonzero momentum, this mixing is not equal to zero. However, it
can be significant only when the LO and TO phonon frequencies are in
resonance. For soft TO phonon modes in strongly polar crystals like
SrTiO$_{3}$, this is not the case, and hence the aforesaid LO-TO phonon mixing
is expected to be of a relatively small importance. When the non-diagonal
blocks of $\mathbb{A}\left(  \omega\right)  $ are neglected, it is reduced to
a quasi-diagonal form of two blocks describing, respectively, Fr\"{o}hlich and
2TO contributions. Without loss of generality, we can choose axes in
coordinate and momentum spaces such that $\mathbf{P}\parallel Oz$, so that
$P_{z}=P,P_{x}=P_{y}=0$. In this basis, the first block is:
\[
\mathbb{A}^{\left(  {\text{LO}}\right)  }\left(  \omega\right)  =\left(
\begin{array}
[c]{ccc}%
A_{xx} & 0 & 0\\
0 & A_{yy} & 0\\
0 & 0 & A_{zz}%
\end{array}
\right)
\]
with the matrix elements%
\begin{equation}
A_{jj}=1-2\hbar\sum_{\mathbf{q},a=1,2,3}\alpha_{\mathbf{q}}^{\left(  a\right)
}k_{j}^{2}.
\end{equation}
Also the second block results in the diagonal matrix:%
\begin{equation}
\mathbb{A}^{\left(  2\mathrm{TO}\right)  }\left(  \omega\right)  =\left(
\begin{array}
[c]{ccc}%
B_{xx} & 0 & 0\\
0 & B_{yy} & 0\\
0 & 0 & B_{zz}%
\end{array}
\right)  \label{second}%
\end{equation}
with the matrix elements:%
\begin{equation}
B_{jj}=1-\frac{g_{2}}{\hbar V}\sum_{\mathbf{q},a=1,2}\lambda_{\mathbf{q}%
}^{\left(  a\right)  }\left(  1-\frac{k_{j}^{2}}{k^{2}}\right)  .
\end{equation}

As the Fr\"{o}hlich polaron self-energy within the approach of squeezed phonon
states is already thoroughly studied in the literature
\cite{Kandemir,Shchadilova,Tulub1962,Tulub2015,Porsch}, we focus on the
contribution for the polaron self-energy for a 2TO interaction. The
determinant of the matrix $\mathbb{A}^{\left(  2\mathrm{TO}\right)  }\left(
\omega\right)  $ is%
\begin{equation}
\det\mathbb{A}^{\left(  2\mathrm{TO}\right)  }\left(  \omega\right)  =%
{\displaystyle\prod\limits_{j=x,y,z}}
\left(  1-\frac{g_{2}}{\hbar V}\sum_{\mathbf{q},a=1,2}\lambda_{\mathbf{q}%
}^{\left(  a\right)  }\left(  1-\frac{k_{j}^{2}}{k^{2}}\right)  \right)  .
\label{detA}%
\end{equation}
The change in the self-energy resulting from the Bogoliubov-Tyablikov
diagonalization is established in the following manner, following the logical
framework outlined in Ref. \cite{Wentzel}. The eigenfrequencies are solutions
to the equation
\begin{equation}
\det\mathbb{M}\left(  \omega\right)  =0.
\end{equation}
The matrix $\mathbb{M}\left(  \omega\right)  $ is diagonalized using the
Bogoliubov-Tyablikov transformation described above. The transformation
(\ref{BT}) is unitary and can be written as $\mathbb{U}_{\text{BT}}F\left(
b^{\dag},b\right)  \mathbb{U}_{\text{BT}}^{-1}$ for any function of phonon
operators $F\left(  b^{\dag},b\right)  $. Hence, the diagonalized matrix is%
\begin{equation}
\mathbb{\tilde{M}}\left(  \omega\right)  =\mathbb{U}_{\text{BT}}%
\mathbb{M}\left(  \omega\right)  \mathbb{U}_{\text{BT}}^{-1}=\left\Vert
\delta_{a^{\prime},a}\delta_{\mathbf{q}^{\prime},\mathbf{q}}\left[  \omega
^{2}-\left(  \nu_{\mathbf{q}}^{\left(  a\right)  }\right)  ^{2}\right]
\right\Vert
\end{equation}
where $\nu_{\mathbf{q}}^{\left(  a\right)  }$ are eigenfrequencies. The
determinant of the matrix $\mathbb{M}\left(  \omega\right)  $ is an invariant
of unitary transformations: $\det\mathbb{\tilde{M}}\left(  \omega\right)
=\det\mathbb{M}\left(  \omega\right)  $. Hence
\begin{equation}
\det\mathbb{M}\left(  \omega\right)  =\prod_{a,\mathbf{q}}\left[  \omega
^{2}-\left(  \nu_{\mathbf{q}}^{\left(  a\right)  }\right)  ^{2}\right]  .
\label{detM1}%
\end{equation}
If the interaction terms in (\ref{Matr}) tend to zero, this determinant turns
to its limiting value
\begin{equation}
\det\mathbb{M}_{0}\left(  \omega\right)  =\prod_{a,\mathbf{q}}\left[
\omega^{2}-\left(  \Omega_{\mathbf{q}}^{\left(  a\right)  }\right)
^{2}\right]  . \label{detM0}%
\end{equation}
Let us introduce the ratio function of $s\equiv\omega^{2}$:%
\begin{equation}
\Delta\left(  s\right)  \equiv\prod_{a,\mathbf{q}}\frac{s-\left(
\nu_{\mathbf{q}}^{\left(  a\right)  }\right)  ^{2}}{s-\left(  \Omega
_{\mathbf{q}}^{\left(  a\right)  }\right)  ^{2}}=\frac{\det\mathbb{M}\left(
\omega\right)  }{\det\mathbb{M}_{0}\left(  \omega\right)  }. \label{Delta}%
\end{equation}
Eigenfrequencies $\nu_{\mathbf{q}}^{\left(  a\right)  }$ are the solutions of
Eq. (\ref{detMeq0}) and also satisfy the equation (\ref{detA}):%
\begin{equation}
\det\mathbb{A}\left(  \omega\right)  =0. \label{eigenf}%
\end{equation}
Consequently, $\det\mathbb{M}\left(  \omega\right)  \propto\det\mathbb{A}%
\left(  \omega\right)  $. Using (\ref{detM0}) and the fact that $\left.
\det\mathbb{A}\left(  \omega\right)  \right\vert _{\left\{  f_{k}\right\}
=0,g_{2}=0}=1$, we find that%
\begin{equation}
\det\mathbb{M}\left(  \omega\right)  =\det\mathbb{M}_{0}\left(  \omega\right)
\det\mathbb{A}\left(  \omega\right)  .
\end{equation}
Consequently, we reproduced here the Wentzel result:%
\begin{equation}
\Delta\left(  \omega^{2}\right)  =\det\mathbb{A}\left(  \omega\right)  .
\end{equation}
The polaron self-energy $\Delta E$ is expressed through the function
$\Delta\left(  s\right)  $ using the identity \cite{Wentzel}:%
\begin{equation}
\frac{\partial}{\partial s}\ln\Delta\left(  s\right)  =\sum_{a,\mathbf{q}%
}\left(  \frac{1}{s-\nu_{\mathbf{q}}^{\left(  a\right)  2}}-\frac{1}%
{s-\Omega_{\mathbf{q}}^{\left(  a\right)  2}}\right)  .
\end{equation}
For any analytic function $F\left(  s\right)  $ this can be expressed via the
Cauchy integral formula as%
\begin{equation}
\sum_{a,\mathbf{q}}\left(  F\left(  \nu_{\mathbf{q}}^{\left(  a\right)
2}\right)  -F\left(  \Omega_{\mathbf{q}}^{\left(  a\right)  2}\right)
\right)  =-\frac{1}{2\pi i}\oint_{C}ds~\frac{\partial F\left(  s\right)
}{\partial s}\ln\Delta\left(  s\right)  ,
\end{equation}
where the contour $C$ embraces all points $s=\nu_{\mathbf{q}}^{\left(
a\right)  2}$ and $s=\Omega_{\mathbf{q}}^{\left(  a\right)  2}$ as shown in
Fig. \ref{fig:Contour}, and the path direction along the contour is counterclockwise.

For the self-energy, $F\left(  s\right)  =\sqrt{s}$. Because the contour $C$
lies in the area where $\operatorname{Re}s>0$, $\sqrt{s}$ is an analytic
single-valued function in that region. Hence
\begin{equation}
\Delta E=-\frac{\hbar}{8\pi i}\oint_{C}ds~\frac{1}{\sqrt{s}}\ln\Delta\left(
s\right)  \label{DE1}%
\end{equation}
where the factor $\Delta\left(  s\right)  $ is related to the matrix
$\mathbb{A}$ as%
\begin{equation}
\Delta\left(  \omega^{2}\right)  =\det\mathbb{A}\left(  \omega\right)  .
\label{W}%
\end{equation}
The particular 2TO contribution to the polaron energy is then%
\begin{align}
\Delta E^{\left(  2\mathrm{TO}\right)  }  &  =-\frac{\hbar}{8\pi i}\oint%
_{C}ds~\frac{1}{\sqrt{s}}\ln\Delta^{\left(  2\mathrm{TO}\right)  }\left(
s\right)  ,\\
\Delta^{\left(  2\mathrm{TO}\right)  }\left(  s\right)   &  =%
{\displaystyle\prod\limits_{j=x,y,z}}
\left(  1-\frac{g_{2}}{\hbar V}\sum_{\mathbf{q},a=1,2}\frac{\varkappa
_{\mathbf{q}}^{2}\Omega_{\mathbf{q}}^{\left(  a\right)  }}{s-\left(
\Omega_{\mathbf{q}}^{\left(  a\right)  }\right)  ^{2}}\left(  1-\frac
{k_{j}^{2}}{k^{2}}\right)  \right)  .
\end{align}
This is equation \eqref{log} from the main text.

\section{Diagrammatic Monte Carlo for the 2TO interaction}

The DiagMC method employed to obtain the polaron energies for the 2TO
interaction {in Fig.~\ref{fig:gsenergy} (\emph{b})} is largely based on the
section \textit{Momentum space representation} in Ref.~\cite{PRBL2023} and
further described in its Supplemental Material.

Consider the Hamiltonian in Eq.~\ref{Ham}. Assuming a symmetric choice for the
polarization vectors such that $\mathbf{e}_{\mathbf{q}}^{(1)}=\mathbf{e}%
_{-\mathbf{q}}^{(1)}$ and $\mathbf{e}_{\mathbf{q}}^{(2)}=\mathbf{e}%
_{-\mathbf{q}}^{(2)}$, $H_{2\mathrm{TO}}$ can be expressed in the form
\begin{equation}
H_{2\mathrm{TO}}=\sum\limits_{\mathbf{k},\mathbf{q,q^{\prime}},a,a^{\prime}%
}V_{\mathbf{q},\mathbf{q}^{\prime}}^{(a,a^{\prime})}\;a_{\mathbf{k}%
+\mathbf{q+q^{\prime}}}^{\dagger}a_{\mathbf{k}}(b_{-\mathbf{q}}^{(a)\dagger
}+b_{\mathbf{q}}^{(a)})(b_{-\mathbf{q^{\prime}}}^{(a^{\prime})\dagger
}+b_{\mathbf{q^{\prime}}}^{(a^{\prime})})
\end{equation}
which is the same as in Ref.~\cite[Suppl. material, Eq.~17]{PRBL2023}. The
constant vertex $(g_{2}\Omega/4)/(2\pi)^{3}$ is replaced by
\begin{equation}
V_{\mathbf{q},\mathbf{q}^{\prime}}^{(a,a^{\prime})}=\frac{1}{(2\pi)^{3}}%
\frac{3\pi^{2}\alpha_{T}}{4}\frac{\mathbf{e}_{\mathbf{q}}^{(a)}\cdot
\mathbf{e}_{\mathbf{q}^{\prime}}^{(a^{\prime})}}{\sqrt{\omega_{\text{TO}%
}^{(a)}(\mathbf{q})\,\omega_{\text{TO}}^{(a^{\prime})}(\mathbf{q^{\prime}})}},
\end{equation}
with the following choice of polarization vectors
\begin{align}
\mathbf{e}_{\mathbf{q}}^{(1)}  &  =\mathbf{e}_{\mathbf{q}}^{\theta}=%
\begin{pmatrix}
\cos(\theta)\cos(\phi), & \cos(\theta)\sin(\phi), & -\sin(\theta)
\end{pmatrix}
\\
\mathbf{e}_{\mathbf{q}}^{(2)}  &  =\mathbf{e}_{\mathbf{q}}^{\phi}=%
\begin{pmatrix}
-\sin(\phi), & \cos(\phi), & 0
\end{pmatrix}
\cdot\text{sgn}(\cos(\theta))
\end{align}
where
\begin{align}
\theta &  =\arccos(q_{z}/q)\\
\phi &  =\arctan(q_{y}/q_{x}).
\end{align}

In order to impose a momentum cutoff, every time a phonon momentum is to be
extracted in the MC updates, it is chosen uniformly inside a sphere in
momentum space with radius $k_{0}$. Consequently, the acceptance ratio of the
\textit{Add/remove 2-loop} and \textit{Add/remove 3-loop} updates must be
modified to contain, respectively, $U(\mathbf{q}_{1}, \mathbf{q}_{2}) = (4/3
\pi k_{0}^{3})^{-2}$ and $U(\mathbf{q}_{1}, \mathbf{q}_{2}, \mathbf{q}_{3}) =
(4/3 \pi k_{0}^{3})^{-3}$. The uniform choice of a particular polarization for
the inserted phonon yields a further probability factor of 1/4 (2-loop) or 1/8
(3-loop) that must be taken into account in the acceptance ratio.

The summation of the complete series of 1-loop diagrams can be analytically
calculated and corresponds precisely to the first order energy correction
(\ref{WC}). As in Ref.~\cite{PRBL2023}, to alleviate the sign problem in the
positive coupling regime, these diagrams can be included into a renormalized
electron propagator with a dispersion shifted by (\ref{WC}).

\end{document}